\documentclass[aps,pre,preprint, showpacs, superscriptaddress]{revtex4}
\usepackage{latexsym}
\usepackage{natbib}
\usepackage{amsmath}
\usepackage{amsfonts}
\usepackage{amssymb}
\usepackage[T1]{fontenc}
\usepackage[dvips]{color,graphicx}   
\usepackage{verbatim}   
\usepackage{t1enc}
\usepackage{anysize}
\usepackage{tocbibind}
\usepackage{subfig}
\usepackage[percent]{overpic} 
\usepackage{float}  
\usepackage{appendix} 
\usepackage{booktabs} 
\usepackage{natbib}
\usepackage{array}
\usepackage[alpine,clock,electronic,geometry,misc,weather]{ifsym}

\begin{document}
\title{Intermittent social distancing strategy for epidemic control}
\author{L. D. Valdez} \affiliation{Instituto de Investigaciones
  F\'isicas de Mar del Plata (IFIMAR)-Departamento de F\'isica,
  Facultad de Ciencias Exactas y Naturales, Universidad Nacional de
  Mar del Plata-CONICET, Funes 3350, (7600) Mar del Plata, Argentina.}
\author{P. A. Macri}
\affiliation{Instituto de Investigaciones F\'isicas de Mar del Plata
  (IFIMAR)-Departamento de F\'isica, Facultad de Ciencias Exactas y
  Naturales, Universidad Nacional de Mar del Plata-CONICET, Funes
  3350, (7600) Mar del Plata, Argentina.}
\author{L. A. Braunstein} \affiliation{Instituto de Investigaciones
  F\'isicas de Mar del Plata (IFIMAR)-Departamento de F\'isica,
  Facultad de Ciencias Exactas y Naturales, Universidad Nacional de
  Mar del Plata-CONICET, Funes 3350, (7600) Mar del Plata, Argentina.}
\affiliation{Center for Polymer Studies, Boston University, Boston,
  Massachusetts 02215, USA}
\begin{abstract}
We study the critical effect of an intermittent social distancing
strategy on the propagation of epidemics in adaptive complex
networks. We characterize the effect of our strategy in the framework
of the susceptible-infected-recovered model. In our model, based
on local information, a susceptible individual interrupts the contact
with an infected individual with a probability $\sigma$ and restores
it after a fixed time $t_{b}$. We find that, depending on the network
topology, in our social distancing strategy there exists a cutoff threshold
$\sigma_{c}$ beyond which the epidemic phase disappears. Our results are
supported by a theoretical framework and extensive simulations of the
model. Furthermore we show that this strategy is very efficient
because it leads to a ``susceptible herd behavior'' that protects a
large fraction of susceptibles individuals. We explain our results using
percolation arguments.
\end{abstract}

\maketitle

\section{Introduction}

The study of the topology of complex networks, and the dynamical
processes that use these networks as substrate to spread, has recently
generated great interest in the scientific community
\cite{Boc_01,Dor_02,Pas_01}. In the past, studies of dynamic
processes such as the spreading of rumors, opinions, and diseases on
static networks, were concentrated on how their topology affects these
processes \cite{Boc_01,Dor_02,Pas_01}. However, it is known that these
processes evolve on top of networks where the topology changes with
time \cite{Gro_01}. As a consequence, recently many researchers began
to study dynamic networks, and the interplay between the dynamic
process and the network dynamics. Those networks in which the topology
changes regardless of the process taking place on top of them are
called evolutive networks, while those networks that change their
topology to mitigate or promote these processes are called adaptive
networks~\cite{Gro_01,Lag_01}. Adaptive networks have been
investigated in many disciplines such as social sciences,
epidemiology, biology, etc.~\cite{Gro_01,Sch_02,Vaz_02}. In these
networks there is a coevolution between the link dynamics and the
state of the nodes which leads to a collective phenomenon on adapting
networks. As an example, in a network of routes, where the nodes are
cities and the links are the routes connecting them, some overloaded
paths become dysfunctional and new paths between cities are built to
avoid the congestion. In the analysis of opinion formation on social
networks, nodes usually tend to rewire or break their links with
individuals with different opinions, leading in some models to a
network fragmentation into components or clusters in which all members
have the same opinion~\cite{Hol_01,Zan_01,Vaz_02}. In biological
networks, such as the vascular system, after an arterial occlusion
collateral vessels grows in order to increase the blood flow to
neighboring tissues~\cite{Wol_01}.  Similarly, in the widely studied
epidemic models on static networks, adaptive processes are used to
model strategies that reduce the impact of the disease
spreading~\cite{Gro_02,Gro_03,Lag_01}.

One of the most popular models in epidemiology that reproduces
seasonal diseases is the susceptible-infected-recovered model
(SIR)~\cite{Boc_01,And_01}, where individuals can be in one of three
states, susceptible ($S$), infected ($I$) or recovered ($R$). In its
classical formulation, an infected node infects a susceptible neighbor
with probability $\beta$ and recovers with a certain fixed
probability, which implies an exponential distribution of times for
which individuals remain infected. However, this distribution is
rarely realistic and for most seasonal diseases it has a sharp peak
around an average value~\cite{KarNew_01}. As a consequence, some
studies~\cite{Lag_02,New_05,Par_01} have used a different version of
the SIR model in which an infected individual recovers after a fixed
time $t_r$, called the recovery time.

 It is well known that in the SIR model on static networks, the size
 of the infection is governed by the effective probability of
 infection or transmissibility $T$ of the disease, where
 $T=1-(1-\beta)^{t_r}$. In turn, it was shown that this model can be
 mapped into a link percolation process~\cite{Gra_01,New_05}, where
 $T$ plays the role of the link occupancy probability $p$ in
 percolation. In a percolation process, there is a critical
 probability $p_{c}$ where the finite cluster size distribution
 $n_{s}$ behaves as $n_{s}\sim s^{-\tau}$ in the thermodynamic
 limit. Above this threshold a ``giant component'' appears. As a
 consequence of the mapping between percolation and the SIR model; in
 the latter there is an epidemic threshold at $T_{c}=p_{c}$ below
 which the disease is an outbreak were the infection reaches a small
 fraction of the population, which is equivalent to having finite
 clusters in percolation, while above $T_{c}$ an epidemic develops
 corresponding to the emergence of a percolating giant
 component~\cite{Mil_02,New_03}. This threshold, in uncorrelated
 static networks, depends only on the degree distribution $P (k)$,
 where $k$ is the degree or the number of links that a node can
 have. In particular, for Erd\"os-R\'enyi (ER) networks,
 $P(k)=e^{-\langle k \rangle}\langle k \rangle^{k}/k!$ where $\langle
 k \rangle$ is the mean connectivity, the threshold is
 $T_{c}=1/\langle k \rangle$. However, in pure scale-free (SF)
 networks $P (k) \sim k ^{-\lambda}$, where $\lambda$ is the broadness
 of the distribution, in the thermodynamic limit $T_c\to 0$ for
 $\lambda <3$, which means that the epidemic spreads for any value of
 $T$. This last result indicates that highly heterogeneous networks,
 such as many theoretical social networks, are very likely to develop
 an epidemic~\cite{New_05,Coh_01}.

While it is well known how the topology affects the SIR process in
static networks, there is very little literature about this model on
adaptive networks. Recently, Lagorio $et\;al.$~\cite{Lag_01} studied
two different strategies to mitigate the spread of a disease in an
adaptive SIR model. In that model a susceptible node disconnects the
link with an infected individual with a rewiring probability $w$, and
creates a link with another susceptible node. The authors found that
there is a phase transition at a critical rewiring threshold $w_{c}$
separating an epidemic from a non-epidemic phase, which can be related
to static link percolation. A feature of this rewiring process is that
links between susceptible individuals can be established independently
of their previous relationship. In that strategy, the nodes have no
memory because two individuals can be connected independently of their
past. Even though this adaptive process could be representative of
casual contacts between individuals such as those generated in public
buildings like shopping, theaters, etc., that strategy will not work
for other types of interaction such as friendship and working
partners, where individuals preserve their closer contacts. Therefore,
if an individual is separated from its closer neighbors, it will tend
to reconnect with them, at some time, more often than with an unknown
individual. On the other hand, recently Wang $et\;al.$~\cite{Wan_01}
and Van Segbroeck $et\;al.$~\cite{Seg_01} proposed strategies to stop
the spread in the SIR model where the susceptible individuals driven
by fear disconnect their links with their neighbors, infected or not,
without creating a new link. This strategy reduces the number of
contacts permanently, which is efficient but also very inconvenient
from an economical point of view.

In this paper we propose a strategy based on ``intermittent social
distancing'' in adaptive networks and study its efficiency in stopping
the spread of diseases in theoretical and real networks. We found
theoretically that in our model there exists a cutoff threshold that
prevents an epidemic phase. Our results are supported by extensive
simulations. Moreover, we found that the intermittent social
distancing strategy is efficient to protect a large susceptible
cluster. The paper is organized as following: in Sec.~\ref{Sec_A_p},
we derive the theoretical transmissibility for our model and show how
our strategy diminishes the epidemic phase. In Sec.~\ref{Sec_Ep_S}, we
show how the epidemic size is reduced with our strategy and the
agreement between our theoretical approach and the simulations. In
Sec.~\ref{Sec_Herd}, we present a study of the ``susceptible herd
behavior'' that we use as a criterion to evaluate the effectiveness of
our strategy. In Sec.~\ref{SecConc}, we present our conclusions.

\section{Analytical approach}\label{Sec_A_p}

We propose a SIR model in an adaptive network where an infected
individual transmits the disease to a susceptible neighbor with
probability $\beta$ and, if he fails, with probability $\sigma$ the
susceptible individual breaks the link with the infected one for a period
$t_b$. Thus, the effective probability of breaking a link is
$(1-\beta)\sigma$. After a time $t_b$ both nodes are reconnected and
the process is repeated until the infected node recovers at a fixed
time $t_r>t_b$, $i.e.$, there is an intermittent connection between a
susceptible node and its infected neighbor. This mimics a behavioral
adaptation of the society to avoid contacts with infected individuals
by imposing a social distancing during one or more periods of duration
$t_{b}$. Notice that in our model a susceptible node breaks its
links using only local information and not global knowledge as provided by
communication media. In our model, the time is increased by 1 after
every infected node tries to infect its neighbors and the updates are
done after each time step. In this process the dynamic
transmissibility $T(\beta,\sigma,t_r,t_b)\equiv T_{\sigma}$ can be
written as

\begin{eqnarray}\label{Ec.Trans}
T_{\sigma}=\sum_{n=1}^{t_r} \beta(1-\beta)^{n-1}(1-\sigma)^{n-1}+ \beta
\sum_{n=t_{b}+2}^{t_r}\phi(n,t_b,\sigma,\beta);
\end{eqnarray}
where $\phi(n,t_b,\sigma,\beta)$ is given by
\begin{equation}\label{eqSumPhi}
\phi(n,t_b,\sigma,\beta)=
\sum_{u=1}^{\left[\frac{n-1}{t_{b}+1}\right]}\binom{n-u\;t_{b}-1}{u}\sigma^{u}(1-\sigma)^{n-1-u(t_{b}+1)}(1-\beta)^{n-1-u\;t_{b}}.
\end{equation}
and $\left[\cdots \right]$ denotes the integer part function.  The
first term of Eq.~(\ref{Ec.Trans}) is the probability that a node in
the $I$ state transmits the disease to any neighbor node in the $S$
state (before it recovers), considering that the link $S$-$I$ has never
been broken. The second term represents the probability for a node in
the $I$ state to transmit the disease to an $S$ node after that pair
has been disconnected $u$ times for a period $t_b$. The binomial
coefficient of the second term takes into account the number of ways
to arrange $u$ intermittent disconnected periods before the
susceptible becomes infected individual at time $n$. Notice that
$ut_b$ is the total times that the pair $S$-$I$ is broken and represent
a temporal social distancing. In Table~\ref{tabla_casos}, we
illustrate the element $n = 8$ of the second term of
Eq.~(\ref{Ec.Trans}) with $t_r = 10$ and $t_b = 2$.

\begin{table}[H]
\caption{Disconnected periods for a pair $S$-$I$ with $t_r=10$
  (recovery time), $t_b=2$ (disconnection period), and $n=8$ (time of
  infection). The first column represents the number of disconnected
  periods $u$ before $n=8$, the second column is a typical
  configuration, the third column is the probability of that
  configuration, and the fourth column is the number of ways to arrange
  $u$ disconnected periods. In the second column, each cell
  corresponds to a unit time. The white cells represent the time unit
  where a link between the $S$ and the $I$ node exists, the gray ones
  correspond to the disconnection period, and in the black cells there
  is no dynamic for the pair $S$-$I$ because the $S$ has been infected
  and now the pair becomes $I$-$I$. Notice that initially the link cannot
  be broken because this disconnection happens only after the $I$
  individual fails to infect the susceptible one, with probability
  $(1-\beta)$. Similarly, two disconnection periods must be separated
  by at least a white cell. When the infection occurs at time $n$, the
  maximum number of disconnected periods is
  $u=\left[(n-1)/(t_b+1)\right]$.}


\renewcommand{\arraystretch}{1}
\begin{tabular}{p{2cm}c@{\hspace{9cm}}c@{\hspace{2cm}}c}  
\toprule[0.05cm]
\multicolumn{1}{c}{$\hspace{1cm}u\hspace{1cm}$}&\multicolumn{1}{c}{Example}&\multicolumn{1}{c}{$\hspace{1.5cm}$Probability$\hspace{1.5cm}$}&\multicolumn{1}{c}{Binomial Coefficient}\\
\cmidrule [0.03cm]{1-4}
\multicolumn{1}{c}{$u=1$}&\multicolumn{1}{m{1cm}} {\includegraphics[scale=0.50]{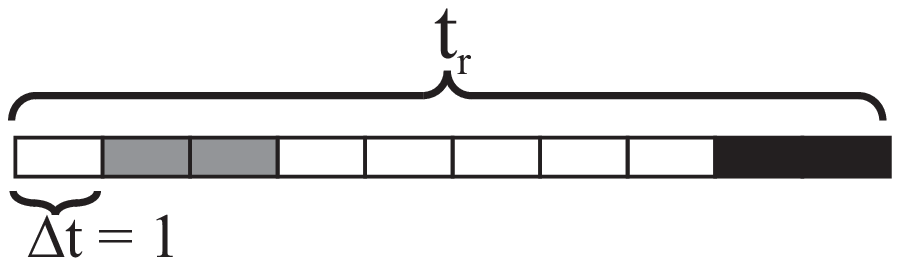}}&\multicolumn{1}{c}{$\beta\;\sigma(1-\sigma)^{4}(1-\beta)^{5}$}&$\binom{8-2-1}{1}=5$\\
\multicolumn{1}{c}{$u=2$}&\multicolumn{1}{c}{\includegraphics[scale=0.50]{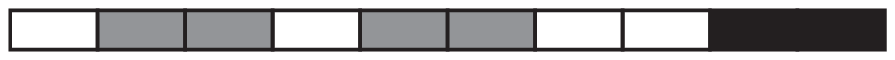}}&\multicolumn{1}{c}{$\beta\;\sigma^{2}(1-\sigma)^{1}(1-\beta)^{3}$}&$\binom{8-4-1}{2}=3$   \\
\bottomrule[0.05cm]
\end{tabular}
\label{tabla_casos}
\end{table}

It is known that the disease becomes an epidemic if the basic
reproductive number $R_{0}\geq 1$, where $R_0$ is the number of
secondary infections. For uncorrelated networks, $R_0$ is related to
the connectivity distribution $P(k)$ through the branching factor
$\kappa$,
\begin{equation}
  R_0 = (\kappa-1) T_{\sigma},
\end{equation}
where $\kappa\equiv \langle k^2\rangle/\langle k \rangle$, and
$\langle k \rangle$ and $\langle k^2 \rangle$ are the first and the
second moments of $P(k)$ respectively. The branching factor is a
measure of the heterogeneity of the network that diverges for SF
networks with $\lambda<3$ in the thermodynamic limit because $\langle
k^2 \rangle\to\infty$~\cite{Coh_01}. Then the critical cutoff
$\sigma_c$, above which the disease dies, as a function of $\beta$,
$t_r$, and $t_b$, can be found through the condition $R_{0}=1$, which
yields
\begin{equation}\label{Critic}
T_{\sigma_{c}}= \frac{1}{\kappa-1}=T_{c},
\end{equation}
\cite{Coh_01,Dun_01} where $T_c$ is the critical transmissibility for
the SIR model in static networks, and consequently our dynamic
  process in the steady state is related to a static topological
  property of the network. This is expected due to the fact that in
  our model the network topology does not change globally in the
  characteristic time scale of the disease spreading, and hence our
  strategy can be understood as an SIR model on a static network but
  with a transmissibility $T_{c}=T_{\sigma_{c}}$.

From Eq.~(\ref{Ec.Trans}) it is straightforward  that in the limit
$\sigma\to 0$, $T_\sigma=T=1-(1-\beta)^{t_r}$. On the other hand, when
$\sigma\to 1$, as the only terms that survive in Eq.~(\ref{eqSumPhi})
are those which fulfills the condition $n-1-u(t_{b}+1)=0$, we obtain
\begin{eqnarray}
T_{\sigma}&=&\beta \left(1+\sum_{u=1}^{\left[\frac{tr-1}{t_{b}+1}\right]}(1-\beta)^{u}\right),\nonumber\\
&=&1-\left(1-\beta\right)^{\left[\frac{t_r-1}{t_b+1}\right]+1}.
\end{eqnarray}

\begin{figure}[H]
\centering
\vspace{0.3cm}
 \includegraphics[scale=0.25]{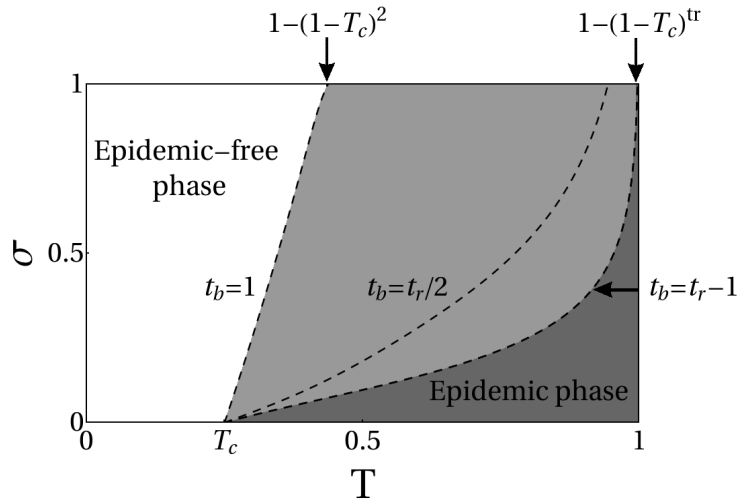}
  \vspace{0.5cm}
\caption{Plot of the epidemic phases in the plane $\sigma-T$ for
  $t_r=20$ and static $T_c=0.25$, where $T$ corresponds to the
  transmissibility in a non adaptive network . The dashed lines
  correspond to the critical dynamic transmissibility $T_{\sigma_c}$
  for (from left to right) $t_b=1$, $t_{b}=t_{r}/2$ and
  $t_b=t_r-1$. For $t_{b}=1$ and $\sigma=1$, the maximum
  transmissibility for the static SIR model for which the epidemic phase is
  $T=1-(1-T_{c})^{t_r/[(t_r+1)/2]}$ or $T\approx 1-(1-T_{c})^{2}$, and the
  $t_r$ dependence disappears.}\label{Fase}
\end{figure}

In Fig.~\ref{Fase}, we plot the plane $\sigma-T$ $\left[T\equiv
T(\sigma=0)\right]$ in order to show that with our strategy the epidemic
phase is reduced compared to the static case.  Notice that for
$t_b=t_r/2$ the epidemic phase shrinks substantially compared to the
case $t_b=1$.
\noindent
In the figure, the light-gray area, delimited between the curves
corresponding to the blocking periods $t_{b}=1$ and $t_{b}=t_{r}-1$,
displays the region of parameters controlled by the intervention
strategy. We can see that even the milder intervention $t_{b}=1$
expands the epidemic-free area compared to the static case. On the
other hand, the stronger intervention $t_{b}=t_{r}-1$ drastically
shrinks the epidemic phase as $\sigma$ increases.  From the figure, we
can also see that for $T_c\neq 0$, as the social distancing increases
the epidemic-free phase increases; meanwhile, for a theoretical SF network with
$\lambda\leq 3$, as $T_c=0$, when $N\to \infty$ there is no
epidemic-free phase for any values of $\sigma$ and $t_{b}$. However,
real networks are finite and they are not generally pure uncorrelated
SF networks which implies that our intermittent social distancing strategy
could be applied in these networks.

\begin{figure}[H]
\centering
\vspace{1.0cm}
  \includegraphics[scale=0.25]{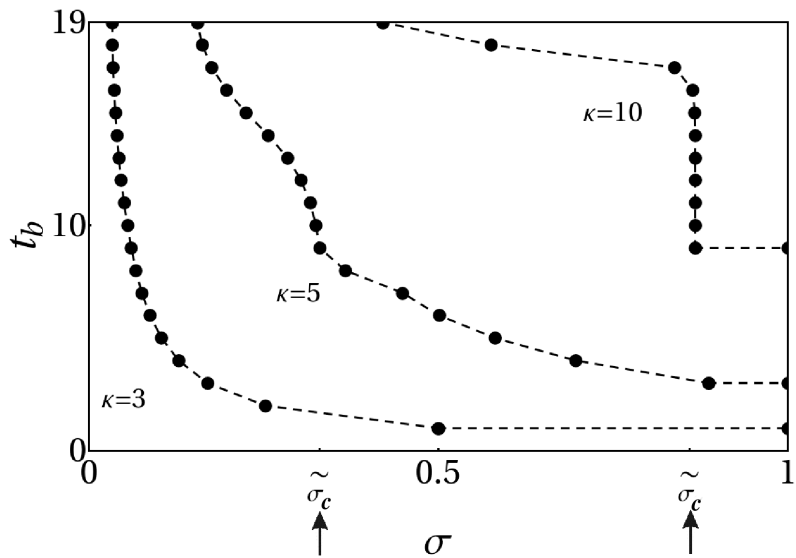}
\caption{Phase diagram for $\sigma$ and $t_{b}$ for $\beta=0.05$ and
  $t_r=20$ for different heterogeneities $\kappa$.  The dashed lines
  with circles represent the interface between the non-epidemic
  (right) and epidemic (left) phases for the different values of
  $\kappa$. Notice that $\widetilde{\sigma_{c}}=0.33$ and
  $\widetilde{\sigma_{c}}=0.86$ correspond to the critical cutoff
  probabilities where in some region the interface is a vertical line
  for $\kappa=5$ and $\kappa=10$, respectively.\label{Phases_SF_ER}}
\end{figure}

In order to determine how the heterogeneity of the network affects the
strategy performance, in Fig.~\ref{Phases_SF_ER} we plot a phase
diagram in the plane $t_b-\sigma$ for fixed $t_r$ and different values
of $\kappa$.

\noindent We can see that as the heterogeneity $\kappa$ increases, the
social distancing $t_{b}$ and $\sigma$ have to increase in order to
prevent the epidemic phase. Surprisingly, in high heterogeneous
networks, we find that for $t_b\geq t_{r}/2$, the critical cutoff
probability $\sigma_c=\widetilde{\sigma_{c}}$ is almost constant (see the
Appendix~\ref{AppenA}) with
\begin{eqnarray}
 \widetilde{\sigma_{c}}&\sim&\frac{\beta}{1-\beta}\left(\frac{1}{T_{c}}-1+\sqrt{\frac{1-T_{c}}{T_{c}^{2}}}\right),\label{Patrick_Star}\\
&\sim& \frac{\beta}{(1-\beta)}\left[\kappa-2+\sqrt{(\kappa-2)(\kappa-1)}\right],\label{SpongeBob}
\end{eqnarray}
$i.e.$, for very heterogeneous networks above $t_{b}=t_{r}/2$,
$\sigma_{c}$ does not change the transmissibility; then the best and
least expensive strategy is to reconnect at $t_b=t_r/2$. This means
that if we know the average duration of a disease $t_{r}$, individuals
can return to their activities with low risk just after half of the
characteristic time. From Eq.~(\ref{SpongeBob}), it is straightforward
that for $\widetilde{\sigma_{c}}=1$ there exists an upper value of
$\kappa \equiv \kappa^{\lim}$ that depends only on $\beta$, where
$\kappa^{\lim}$ is given by
\begin{eqnarray}\label{kappalim}
\kappa^{\lim}&=&\frac{2-(1-\beta)^2}{1-(1-\beta)^{2}},
\end{eqnarray}
where this strategy can be applied.  From Eq.~(\ref{Patrick_Star}) the
limit $\widetilde{\sigma_{c}}=1$ can also be expressed in terms of the
minimum critical transmissibility,
\begin{eqnarray}\label{TClim}
T_{\sigma_{c}}^{\lim}&=&2\beta-\beta^{2}.
\end{eqnarray}
As a consequence, for very low infection rates, our strategy predicts
an epidemic-free phase even for highly heterogeneous networks with a
finite epidemic threshold. Notice that since real networks have degree
correlations and clustering, then the relation $T_{c}=1/(\kappa-1)$
does not hold and the only magnitude that matters is $T$, which has to
be measured by the peak of the second giant component~\cite{Wu_01}. In
those cases, for $\widetilde{\sigma_{c}}=1$ we have to use
Eq.~(\ref{TClim}) instead of Eq.~(\ref{kappalim}). We found that for
the condensed matter coauthorship network ~\cite{New_04} ($T_c=0.026$)
and mathematics coauthorship network~\cite{Pal_01} ($T_{c}=0.050$) the
epidemic spread can be stopped in diseases with $\beta\leq0.013$ and
$\beta\leq 0.025$, respectively. Similar values of $\beta$ were used
for real networks by Kitsak $et\;al.$ \cite{Kit_01}.

\section{Numerical results}\label{Sec_Ep_S}

In our model at the initial stage, all nodes are susceptible and we
infect a randomly chosen node (patient zero) in the biggest component
of our network. From the patient zero the disease spreads to its
neighbors according to the rules of our model described above: an
infected individual transmits the disease to a susceptible neighbor
with probability $\beta$ and, if it fails, with probability $\sigma$
the susceptible individual breaks the link with the infected one for a period
$t_b$. After a time $t_b$ both nodes are reconnected and the process
is repeated until the infected node recovers at a fixed time
$t_r>t_b$. The time is increased by 1 after every
infected node tries to infect its neighbors and the updates are done
after each time step.

 All our results are presented for $t_{r}=20$ but qualitatively all
 the results are the same for $t_{r}>1$.

\begin{figure}[H]
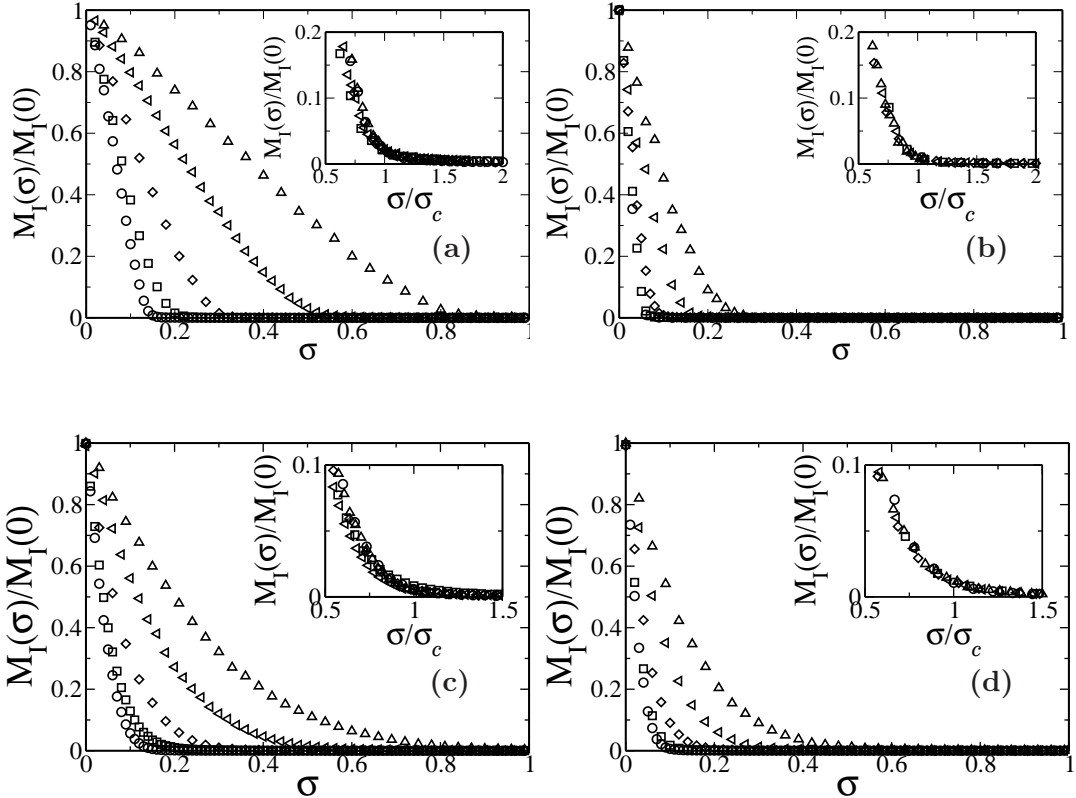

\centering
\vspace{1.0cm}
  \begin{overpic}[scale=0.27]{Fig5.eps}
    \put(80,20){{\bf{(a)}}}
  \end{overpic}\vspace{0.5cm}
  \begin{overpic}[scale=0.27]{Fig6.eps}
    \put(80,20){\bf{(b)}}
  \end{overpic}\vspace{0.5cm}
  \begin{overpic}[scale=0.27]{Fig7.eps}
    \put(80,20){{\bf{(c)}}}
  \end{overpic}\vspace{0.5cm}
  \begin{overpic}[scale=0.27]{Fig8.eps}
    \put(80,20){\bf{(d)}}
  \end{overpic}\vspace{0.5cm}
\caption{$M_{I}(\sigma,t_{b})/M_{I}(\sigma=0)$ vs. $\sigma$ in an ER
  network with $\langle k \rangle =4$, $N=10^{4}$, $t_r=20$, for
  different values of $t_b$: $t_b=3$~($\triangle$),
  $t_b=5$~(\textnarrow{\BigTriangleLeft} ), $t_{b}=10$~(\BigDiamondshape),
  $t_{b}=15$~($\square$) and $t_{b}=19$~($\bigcirc$) for $\beta=0.05$
  with original transmissibility $T=0.64$ (a) and $\beta=0.025$ with
  $T=0.40$ (b); $M_{I}(\sigma,t_{b})/M_{I}(\sigma=0)$ vs. $\sigma$ in
  a SF network with $\lambda=3.5$ and minimal connectivity $k_{min}=2$
  for $t_{r}=20$, $\beta=0.075$ with original transmissibility
  $T=0.79$(c) and $\beta=0.05$ with original transmissibility $T=0.64$
  (d). In the insets we show an enlargement of the main plot, rescaled
  in the abscissa by the factor $\sigma_{c}$, obtained from
  Eq.~(\ref{Ec.Trans}). Our simulations were averaged over $10^4$
  realizations.}\label{Ni_ER}
\end{figure}
In Fig.~\ref{Ni_ER}, we plot the relative epidemic size
$M_{I}/M_{I}(0)\equiv M_{I}(\sigma;t_{b})/M_{I}(\sigma=0)$ as a
function of $\sigma$ for ER and SF
  networks for different values of $t_{b}$
and $t_r=20$. From the plot we can see that $M_{I}(\sigma)$ decreases
as $\sigma$ and $t_{b}$ increase compared to the static case
$M_{I}(0)$. We can also see that a critical probability $\sigma_{c}$
exists, which can be obtained theoretically from Eq.~(\ref{Critic}),
above which the disease dies. Then, depending on how virulent is the
disease without intervention, we can control $t_{b}$ in order to stop
the spread. In the insets of the figures, we collapse all the curves
using the value of $\sigma_{c}$ obtained from Eq.~(\ref{Critic}). The
collapse close to the critical value shows the excellent agreement
between the theoretical value of $\sigma_{c}$ and the simulations.

\section{Susceptible herd behavior}\label{Sec_Herd}

With our intermittent social distancing strategy, the susceptible
nodes dynamically reduce their contact with the nodes in the infected
cluster, mitigating the spread of the disease, $i.e.$, our strategy
produces a resistance to the disease which we call ``susceptible herd
behavior''. As a result of the coevolutive process, at the end of the
spreading there is only one cluster composed of recovered individuals
and, depending on $T_{\sigma}$, one or more susceptible clusters that
give raise to a cluster size distribution of susceptible individuals
or ``voids'' \cite{Exxon_05}. In our model, the cluster size
distribution of voids or susceptible individuals is important since
the formation of a susceptible herd behavior induced by the network
dynamics also measures how effective our strategy is to preserve a
whole part of the society safe from the disease. Next, we derive the
value of the transmissibility for which a susceptible crowding
develops, $i.e.$, the value below which our strategy is efficient.

Describing the growth of an epidemic cluster as a Leath
process~\cite{Lea_01,Bra_01} for a value of the link occupancy probability
$p\equiv T_{\sigma}$ and denoting by $f_{n}(p)$ the probability that a cluster
reaches the $nth$ generation following a link, then the probability
$f_{\infty}(p)$ that a link leads to a giant component when $n\to
\infty$ is given by
\begin{equation}\label{it03}
f_{\infty}(p)=\sum_{k=1}\frac{kP(k)}{\langle k\rangle} \left[ 1-p\;f_{\infty}(p)\right]^{k-1},
\end{equation}
where $f_{\infty}(p)$ is the solution of
\begin{equation}\label{eqpinf}
f_{\infty}(p)=1-G_{1}(1-p\;f_{\infty}(p)),
\end{equation}
and $G_{1}(x)=\sum_{k=1}^{\infty}{kP(k)}/{\langle k\rangle}x^{k-1}$.

When the ``epidemic'' cluster grows, the size of the void clusters is
reduced as in a node dilution process, since when a link is occupied
a void cluster loses a node and all its edges. Then $f_{\infty}(p)$ is
the probability that a void cluster loses a node. If we denote by
$1-p^{v}$ the fraction of void nodes removed the following relation
holds~\cite{Exxon_04},
\begin{eqnarray}
1-p^{v}=f_{\infty}(p).
\end{eqnarray}
For node percolation, it is known that~\cite{Sta_01}
$P^{v}_{\infty}(p^{v})+\sum s\;n_{s}^{v}=p^{v}$, where
$P^{v}_{\infty}(p^v)$ is the fraction of nodes in the giant void
component and $n_{s}^v$ the number of finite void clusters of size
$s$. The fraction of remaining nodes, below which the giant void
cluster is destroyed corresponds to the critical probability of node
void percolation $p^{v}=p_c^{v}$. At this value $P_{\infty}^{v}(p_{c}^{v})=0$;
then $\sum s\;n_{s}^{v}=p_c^{v}$. This means that at the void transition,
only a fraction $p_{c}^{v}$ of the nodes belong to void
clusters. As a consequence, the fraction of links
$p^{*}$ needed to reach this point fulfills
\begin{equation}\label{condition_p}
p_{c}^{v}=1-f_{\infty}(p^{*}).
\end{equation}
Therefore, from Eqs.~(\ref{eqpinf})~and~(\ref{condition_p})
\begin{equation}\label{pestrellaFinal}
p_{c}^{v}=G_{1}(1-p^{*}(1-p_{c}^{v})),
\end{equation}
where $p_{c}^{v}=G_{1}\left((G_{1}^{'})^{-1}(1)\right)$ and $p^{*}$ is
the solution of Eq.~(\ref{pestrellaFinal}). Notice that at the void
transition $G_{1}^{'}(1-p^{*}+p_{c}^{v}p^{*})=1$. A similar result was
also found by Newman for a process of spreading of two
pathogens~\cite{New_06}, employing a different approach. Using the
mapping between our adaptive SIR model and link percolation, the
transmissibility $T_{c}^{*}$ needed to create a giant component of
crowded susceptible individuals is $T_{c}^{*}=p^{*}$. Thus, for a
disease spreading in an ER network, the dynamical critical
transmissibility for the giant susceptible cluster is given by
$T_{\sigma}^{*}=-T_{\sigma_{c}}\ln(T_{\sigma_{c}})/(1-T_{\sigma_{c}})$.

\begin{figure}[H]
\centering
\vspace{1cm}
 \includegraphics[scale=0.35]{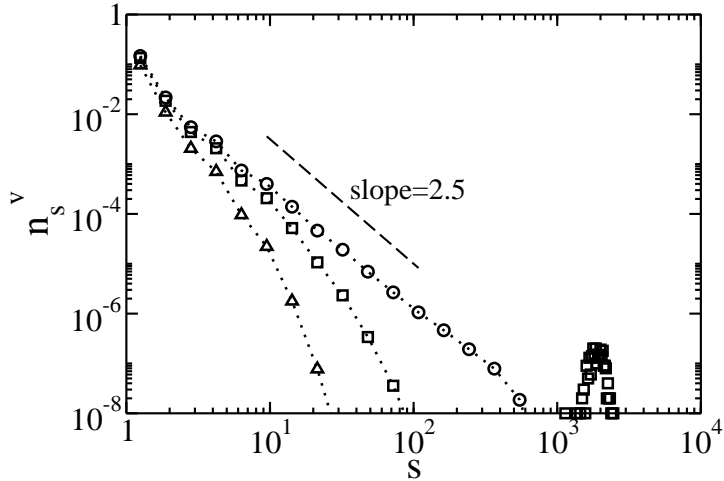}
\caption{Log-log plot of cluster size distribution of susceptible
  individuals for an ER network with $\langle k \rangle=4$
  ($T^{*}_{\sigma}=0.46$) and $N=10^4$ at: $T_{\sigma}=0.64$
  ($\triangle$), $T_{\sigma}=0.46$ ($\bigcirc$) and $T_{\sigma}=0.40$
  ($\square$). The dotted lines are a guide to the eye and the dashed
  line is the result of a power law fitting for
  $T_{\sigma}^{*}$. Notice that we add in $n_{s}^{v}$ the giant
  component. \label{nsErd}}
\end{figure}

In Fig.~\ref{nsErd}, we compare the cluster size distribution of
susceptible individuals or voids $n_{s}^{v}$ for different values of the
transmissibility where we distinguish outbreaks from
epidemics~\cite{Lag_02,Ken_01,New_05} on an ER network with $\langle k
\rangle=4$.

\noindent For $T_{\sigma}>T_{\sigma}^{*}$ we obtain only small
clusters of susceptible nodes that decay faster than exponentially. On
the other hand, for $T_{\sigma}<T^{*}_{\sigma}$ a giant susceptible
cluster appears, meaning that our strategy is efficient producing
large connected clusters of susceptible nodes that crowd in order to
protect themselves. When with our strategy the transmissibility is
reduced to $T_{\sigma}=T^{*}_{\sigma}$ the epidemic spreading slows
down and the distribution $n_{s}^{v}$ decays as a power law with the
same exponent $\tau=5/2$ as in a void node percolation transition, in
contrast with the distribution of infected sizes of outbreaks which at
criticality goes as $ n_{S}^{I}\sim s^{- \tau +1} $ as in a Leath
process~\cite{Lea_01}. These results confirms the importance of the
interplay between void nodes and link percolation in our model.  .

\section{Conclusions}\label{SecConc}

In this paper we present a novel adaptive strategy based on
intermittent social distancing in an adaptive SIR model. In the
intermittent social distancing strategy a susceptible individual
breaks the link with the infected neighbor with a cutoff probability
$\sigma$ and then both individuals are reconnected after a time
$t_{b}$ before the infected individual recovers at $t_{r}$.  Using the
framework of percolation theory, we derive the dynamical
transmissibility, and we find that there exists a critical cutoff
$\sigma_{c}$ where the epidemic spread is stopped for non-highly
heterogeneous networks. We show that in some real networks our
intermittent social distance strategy could stop the epidemic
spreading for not very virulent diseases. We find an excellent
agreement between the theory and the simulations. For heterogeneous
networks, we find that a very high value of $t_{b}$ does not lead to a
decrease of the epidemic, which implies that the less expensive
strategy is to chose $t_b=t_r/2$. Finally we verify that our strategy
reduces the transmissibility below a value where a susceptible
crowding is produced. Any clever strategy used to reduce the disease
from spreading should protect the population at the least economic
cost. We believe that our strategy, which allows us to control the
disconnection of periods through $t_{b}$, is a very convenient strategy
because it creates a susceptible herd cluster. Our present findings
could be used as a support and reference guidance for the development
of further strategies to stop diseases in real networks.

\acknowledgements This work was supported by UNMdP and FONCyT (Pict
Grant No.  0293/2008).  The authors thank Federico Vazquez and Camila
Buono for useful discussions.

\appendix
\section{}\label{AppenA}
For $t_b\geq \left[(tr-1)/2\right]\approx t_{r}/2$ and
$t_{b}<t_{r}-1$, in Eq.~(\ref{eqSumPhi}) the summation has only
one term, because for $t_b\gtrsim t_{r}/2$ we have only one period of
disconnection, $i.e.$, $[(n-1)/(t_{b}+1)]=1$. Intuitively, if
$t_b\gtrsim t_{r}/2$ then in order to compute the transmissibility, we
have to consider at most only one break period since, otherwise the
transmission of the disease is not possible. Denoting by
$\delta=(1-\sigma)(1-\beta)$, then Eq.~(\ref{Ec.Trans}) reduces to
\begin{eqnarray}
T_{\sigma}&=&\sum_{n=1}^{t_r}
\beta\;\delta^{n-1}+\beta(1-\beta)\sigma\sum_{n=t_{b}+2}^{t_r}(n-t_{b}-1)\delta^{n-2-t_{b}}\nonumber \\
&=&\frac{\beta\left(1-\delta^{t_{r}}\right)}{1-\delta}+\frac{\beta\sigma}{(1-\sigma)}\frac{\delta}{(1-\delta)^2}\left[1+\delta^{t_r-t_{b}-1}\left(\delta(t_{r}-t_{b}-1)+t_{b}-t_r\right)\right].
\end{eqnarray}
Neglecting the higher powers of $\delta$ (which hold for
$\sigma\to1$, $\beta\to 1$, or high values of $t_r-t_b$), we obtain
\begin{eqnarray}
T_{\sigma}=\frac{\beta}{1-\delta}+\frac{\beta\sigma(1-\beta)}{(1-\delta)^{2}}.
\end{eqnarray}
Notice that $T_{\sigma}$ loses all the dependence on $t_{r}$ and
$t_{b}$ as shown in Fig.~\ref{Phases_SF_ER}. With this approximation
and using the fact that $T_{\sigma_c}=T_{c}=1/(1-\kappa)$, the
critical cutoff $\sigma_{c}=\widetilde{\sigma_{c}}$ is given by
\begin{eqnarray}
\widetilde{\sigma_{c}}=\frac{\beta}{1-\beta}\left(\frac{1}{T_{c}}-1+\sqrt{\frac{1-T_{c}}{T_{c}^{2}}}\right).
\end{eqnarray}
If Eq.~(\ref{Critic}) holds, then
\begin{eqnarray}
\widetilde{\sigma_{c}}=\frac{\beta}{1-\beta}\left[\kappa-2+\sqrt{(\kappa-1)(\kappa-2)}\right].
\end{eqnarray}

\bibliography{VMBbib.bib}

\begin{thebibliography}{35}
\expandafter\ifx\csname natexlab\endcsname\relax\def\natexlab#1{#1}\fi
\expandafter\ifx\csname bibnamefont\endcsname\relax
  \def\bibnamefont#1{#1}\fi
\expandafter\ifx\csname bibfnamefont\endcsname\relax
  \def\bibfnamefont#1{#1}\fi
\expandafter\ifx\csname citenamefont\endcsname\relax
  \def\citenamefont#1{#1}\fi
\expandafter\ifx\csname url\endcsname\relax
  \def\url#1{\texttt{#1}}\fi
\expandafter\ifx\csname urlprefix\endcsname\relax\def\urlprefix{URL }\fi
\providecommand{\bibinfo}[2]{#2}
\providecommand{\eprint}[2][]{\url{#2}}

\bibitem[{\citenamefont{Boccaletti et~al.}(2006)\citenamefont{Boccaletti,
  Latora, Moreno, Chavez, and Hwang}}]{Boc_01}
\bibinfo{author}{\bibfnamefont{S.}~\bibnamefont{Boccaletti}},
  \bibinfo{author}{\bibfnamefont{V.}~\bibnamefont{Latora}},
  \bibinfo{author}{\bibfnamefont{Y.}~\bibnamefont{Moreno}},
  \bibinfo{author}{\bibfnamefont{M.}~\bibnamefont{Chavez}}, \bibnamefont{and}
  \bibinfo{author}{\bibfnamefont{D.}~\bibnamefont{Hwang}},
  \bibinfo{journal}{Physics Reports} \textbf{\bibinfo{volume}{424}},
  \bibinfo{pages}{175} (\bibinfo{year}{2006}).

\bibitem[{\citenamefont{Dorogovtsev and Mendes}(2003)}]{Dor_02}
\bibinfo{author}{\bibfnamefont{S.~N.} \bibnamefont{Dorogovtsev}}
  \bibnamefont{and} \bibinfo{author}{\bibfnamefont{J.~F.~F.}
  \bibnamefont{Mendes}}, \emph{\bibinfo{title}{Evolution of Networks}}
  (\bibinfo{publisher}{Oxford}, \bibinfo{year}{2003}).

\bibitem[{\citenamefont{Romualdo Pastor~Satorras}(2004)}]{Pas_01}
\bibinfo{author}{\bibfnamefont{A.~V.} \bibnamefont{Romualdo Pastor~Satorras}},
  \emph{\bibinfo{title}{Evolution and Structure of the Internet: a statistical
  approach}} (\bibinfo{publisher}{Cambridge}, \bibinfo{year}{2004}).

\bibitem[{\citenamefont{Thilo~Gross}(2009)}]{Gro_01}
\bibinfo{author}{\bibfnamefont{H.~S.} \bibnamefont{Thilo~Gross}},
  \emph{\bibinfo{title}{Adaptive Networks: Theory, Models and Applications}}
  (\bibinfo{publisher}{Springer}, \bibinfo{year}{2009}).

\bibitem[{\citenamefont{Lagorio et~al.}(2011)\citenamefont{Lagorio, Dickison,
  Vazquez, Braunstein, Macri, Migueles, Havlin, and Stanley}}]{Lag_01}
\bibinfo{author}{\bibfnamefont{C.}~\bibnamefont{Lagorio}},
  \bibinfo{author}{\bibfnamefont{M.}~\bibnamefont{Dickison}},
  \bibinfo{author}{\bibfnamefont{F.}~\bibnamefont{Vazquez}},
  \bibinfo{author}{\bibfnamefont{L.~A.} \bibnamefont{Braunstein}},
  \bibinfo{author}{\bibfnamefont{P.~A.} \bibnamefont{Macri}},
  \bibinfo{author}{\bibfnamefont{M.~V.} \bibnamefont{Migueles}},
  \bibinfo{author}{\bibfnamefont{S.}~\bibnamefont{Havlin}}, \bibnamefont{and}
  \bibinfo{author}{\bibfnamefont{H.~E.} \bibnamefont{Stanley}},
  \bibinfo{journal}{Phys. Rev. E} \textbf{\bibinfo{volume}{83}},
  \bibinfo{pages}{026102} (\bibinfo{year}{2011}).

\bibitem[{\citenamefont{Schwartz and Shaw}(2010)}]{Sch_02}
\bibinfo{author}{\bibfnamefont{I.~B.} \bibnamefont{Schwartz}} \bibnamefont{and}
  \bibinfo{author}{\bibfnamefont{L.~B.} \bibnamefont{Shaw}},
  \bibinfo{journal}{Physics} \textbf{\bibinfo{volume}{3}}, \bibinfo{pages}{17}
  (\bibinfo{year}{2010}).

\bibitem[{\citenamefont{Vazquez et~al.}(2008)\citenamefont{Vazquez, Egu\'iluz,
  and Miguel}}]{Vaz_02}
\bibinfo{author}{\bibfnamefont{F.}~\bibnamefont{Vazquez}},
  \bibinfo{author}{\bibfnamefont{V.~M.} \bibnamefont{Egu\'iluz}},
  \bibnamefont{and} \bibinfo{author}{\bibfnamefont{M.~S.}
  \bibnamefont{Miguel}}, \bibinfo{journal}{Phys. Rev. Lett.}
  \textbf{\bibinfo{volume}{100}}, \bibinfo{pages}{108702}
  (\bibinfo{year}{2008}).

\bibitem[{\citenamefont{Holme and Newman}(2006)}]{Hol_01}
\bibinfo{author}{\bibfnamefont{P.}~\bibnamefont{Holme}} \bibnamefont{and}
  \bibinfo{author}{\bibfnamefont{M.~E.~J.} \bibnamefont{Newman}},
  \bibinfo{journal}{Phys. Rev. E} \textbf{\bibinfo{volume}{74}},
  \bibinfo{pages}{056108} (\bibinfo{year}{2006}).

\bibitem[{\citenamefont{Gil and Zanette}(2006)}]{Zan_01}
\bibinfo{author}{\bibfnamefont{S.}~\bibnamefont{Gil}} \bibnamefont{and}
  \bibinfo{author}{\bibfnamefont{D.~H.} \bibnamefont{Zanette}},
  \bibinfo{journal}{Phys. Lett. A} \textbf{\bibinfo{volume}{356}},
  \bibinfo{pages}{89} (\bibinfo{year}{2006}).

\bibitem[{\citenamefont{Schaper and Scholz}(2003)}]{Wol_01}
\bibinfo{author}{\bibfnamefont{W.}~\bibnamefont{Schaper}} \bibnamefont{and}
  \bibinfo{author}{\bibfnamefont{D.}~\bibnamefont{Scholz}},
  \bibinfo{journal}{Thromb. Vasc. Biol.} \textbf{\bibinfo{volume}{23}},
  \bibinfo{pages}{1143} (\bibinfo{year}{2003}).

\bibitem[{\citenamefont{Gross et~al.}(2006)\citenamefont{Gross, D'Lima, and
  Blasius}}]{Gro_02}
\bibinfo{author}{\bibfnamefont{T.}~\bibnamefont{Gross}},
  \bibinfo{author}{\bibfnamefont{C.~J.~D.} \bibnamefont{D'Lima}},
  \bibnamefont{and} \bibinfo{author}{\bibfnamefont{B.}~\bibnamefont{Blasius}},
  \bibinfo{journal}{Phys. Rev. Lett.} \textbf{\bibinfo{volume}{96}},
  \bibinfo{pages}{208701} (\bibinfo{year}{2006}).

\bibitem[{\citenamefont{Gross and Kevrekidis}(2008)}]{Gro_03}
\bibinfo{author}{\bibfnamefont{T.}~\bibnamefont{Gross}} \bibnamefont{and}
  \bibinfo{author}{\bibfnamefont{I.~G.} \bibnamefont{Kevrekidis}},
  \bibinfo{journal}{EPL (Europhysics Letters)} \textbf{\bibinfo{volume}{82}},
  \bibinfo{pages}{38004} (\bibinfo{year}{2008}).

\bibitem[{\citenamefont{Anderson and May}(1992)}]{And_01}
\bibinfo{author}{\bibfnamefont{R.~M.} \bibnamefont{Anderson}} \bibnamefont{and}
  \bibinfo{author}{\bibfnamefont{R.~M.} \bibnamefont{May}},
  \emph{\bibinfo{title}{Infectious Diseases of Humans: Dynamics and Control}}
  (\bibinfo{publisher}{Oxford University Press, Oxford}, \bibinfo{year}{1992}).

\bibitem[{\citenamefont{Karrer and Newman}(2010)}]{KarNew_01}
\bibinfo{author}{\bibfnamefont{B.}~\bibnamefont{Karrer}} \bibnamefont{and}
  \bibinfo{author}{\bibfnamefont{M.~E.~J.} \bibnamefont{Newman}},
  \bibinfo{journal}{Phys. Rev. E} \textbf{\bibinfo{volume}{82}},
  \bibinfo{pages}{016101} (\bibinfo{year}{2010}).

\bibitem[{\citenamefont{Lagorio et~al.}(2009)\citenamefont{Lagorio, Migueles,
  Braunstein, L\'opez, and Macri}}]{Lag_02}
\bibinfo{author}{\bibfnamefont{C.}~\bibnamefont{Lagorio}},
  \bibinfo{author}{\bibfnamefont{M.}~\bibnamefont{Migueles}},
  \bibinfo{author}{\bibfnamefont{L.}~\bibnamefont{Braunstein}},
  \bibinfo{author}{\bibfnamefont{E.}~\bibnamefont{L\'opez}}, \bibnamefont{and}
  \bibinfo{author}{\bibfnamefont{P.}~\bibnamefont{Macri}},
  \bibinfo{journal}{Physica A: Statistical Mechanics and its Applications}
  \textbf{\bibinfo{volume}{388}}, \bibinfo{pages}{755 } (\bibinfo{year}{2009}).

\bibitem[{\citenamefont{Newman}(2002)}]{New_05}
\bibinfo{author}{\bibfnamefont{M.~E.~J.} \bibnamefont{Newman}},
  \bibinfo{journal}{Physical Review E} \textbf{\bibinfo{volume}{66}},
  \bibinfo{pages}{016128} (\bibinfo{year}{2002}).

\bibitem[{\citenamefont{Parshani et~al.}(2010)\citenamefont{Parshani, Carmi,
  and Havlin}}]{Par_01}
\bibinfo{author}{\bibfnamefont{R.}~\bibnamefont{Parshani}},
  \bibinfo{author}{\bibfnamefont{S.}~\bibnamefont{Carmi}}, \bibnamefont{and}
  \bibinfo{author}{\bibfnamefont{S.}~\bibnamefont{Havlin}},
  \bibinfo{journal}{Phys. Rev. Lett.} \textbf{\bibinfo{volume}{104}},
  \bibinfo{pages}{258701} (\bibinfo{year}{2010}).

\bibitem[{\citenamefont{Grassberger}(1983)}]{Gra_01}
\bibinfo{author}{\bibfnamefont{P.}~\bibnamefont{Grassberger}},
  \bibinfo{journal}{Math. Biosci.} \textbf{\bibinfo{volume}{63}},
  \bibinfo{pages}{157} (\bibinfo{year}{1983}).

\bibitem[{\citenamefont{Miller}(2007)}]{Mil_02}
\bibinfo{author}{\bibfnamefont{J.~C.} \bibnamefont{Miller}},
  \bibinfo{journal}{Phys. Rev. E} \textbf{\bibinfo{volume}{76}},
  \bibinfo{pages}{010101} (\bibinfo{year}{2007}).

\bibitem[{\citenamefont{Newman et~al.}(2001)\citenamefont{Newman, Strogatz, and
  Watts}}]{New_03}
\bibinfo{author}{\bibfnamefont{M.~E.~J.} \bibnamefont{Newman}},
  \bibinfo{author}{\bibfnamefont{S.~H.} \bibnamefont{Strogatz}},
  \bibnamefont{and} \bibinfo{author}{\bibfnamefont{D.~J.} \bibnamefont{Watts}},
  \bibinfo{journal}{Phys. Rev. E} \textbf{\bibinfo{volume}{64}},
  \bibinfo{pages}{026118} (\bibinfo{year}{2001}).

\bibitem[{\citenamefont{Cohen et~al.}(2000)\citenamefont{Cohen, Erez, ben
  Avraham, and Havlin}}]{Coh_01}
\bibinfo{author}{\bibfnamefont{R.}~\bibnamefont{Cohen}},
  \bibinfo{author}{\bibfnamefont{K.}~\bibnamefont{Erez}},
  \bibinfo{author}{\bibfnamefont{D.}~\bibnamefont{ben Avraham}},
  \bibnamefont{and} \bibinfo{author}{\bibfnamefont{S.}~\bibnamefont{Havlin}},
  \bibinfo{journal}{Phys. Rev. Lett.} \textbf{\bibinfo{volume}{85}},
  \bibinfo{pages}{4626} (\bibinfo{year}{2000}).

\bibitem[{\citenamefont{Wang et~al.}(2011)\citenamefont{Wang, Xiao, Wong, Fu,
  Ma, and Cheng}}]{Wan_01}
\bibinfo{author}{\bibfnamefont{Y.}~\bibnamefont{Wang}},
  \bibinfo{author}{\bibfnamefont{G.}~\bibnamefont{Xiao}},
  \bibinfo{author}{\bibfnamefont{L.}~\bibnamefont{Wong}},
  \bibinfo{author}{\bibfnamefont{X.}~\bibnamefont{Fu}},
  \bibinfo{author}{\bibfnamefont{S.}~\bibnamefont{Ma}}, \bibnamefont{and}
  \bibinfo{author}{\bibfnamefont{T.~H.} \bibnamefont{Cheng}},
  \bibinfo{journal}{J. Phys. A: Math. Theor.} \textbf{\bibinfo{volume}{44}},
  \bibinfo{pages}{355101} (\bibinfo{year}{2011}).

\bibitem[{\citenamefont{Van~Segbroeck et~al.}(2010)\citenamefont{Van~Segbroeck,
  Santos, and Pacheco}}]{Seg_01}
\bibinfo{author}{\bibfnamefont{S.}~\bibnamefont{Van~Segbroeck}},
  \bibinfo{author}{\bibfnamefont{F.~C.} \bibnamefont{Santos}},
  \bibnamefont{and} \bibinfo{author}{\bibfnamefont{J.~M.}
  \bibnamefont{Pacheco}}, \bibinfo{journal}{PLoS Comput Biol}
  \textbf{\bibinfo{volume}{6}}, \bibinfo{pages}{e1000895}
  (\bibinfo{year}{2010}).

\bibitem[{\citenamefont{Callaway et~al.}(2000)\citenamefont{Callaway, Newman,
  Strogatz, and Watts}}]{Dun_01}
\bibinfo{author}{\bibfnamefont{D.~S.} \bibnamefont{Callaway}},
  \bibinfo{author}{\bibfnamefont{M.~E.~J.} \bibnamefont{Newman}},
  \bibinfo{author}{\bibfnamefont{S.~H.} \bibnamefont{Strogatz}},
  \bibnamefont{and} \bibinfo{author}{\bibfnamefont{D.~J.} \bibnamefont{Watts}},
  \bibinfo{journal}{Phys. Rev. Lett.} \textbf{\bibinfo{volume}{85}},
  \bibinfo{pages}{5468} (\bibinfo{year}{2000}).

\bibitem[{\citenamefont{Wu et~al.}(2007)\citenamefont{Wu, Lagorio, Braunstein,
  Cohen, Havlin, and Stanley}}]{Wu_01}
\bibinfo{author}{\bibfnamefont{Z.}~\bibnamefont{Wu}},
  \bibinfo{author}{\bibfnamefont{C.}~\bibnamefont{Lagorio}},
  \bibinfo{author}{\bibfnamefont{L.~A.} \bibnamefont{Braunstein}},
  \bibinfo{author}{\bibfnamefont{R.}~\bibnamefont{Cohen}},
  \bibinfo{author}{\bibfnamefont{S.}~\bibnamefont{Havlin}}, \bibnamefont{and}
  \bibinfo{author}{\bibfnamefont{H.~E.} \bibnamefont{Stanley}},
  \bibinfo{journal}{Physical Review E} \textbf{\bibinfo{volume}{75}},
  \bibinfo{pages}{066110} (\bibinfo{year}{2007}).

\bibitem[{\citenamefont{Newman}(2001)}]{New_04}
\bibinfo{author}{\bibfnamefont{M.~E.~J.} \bibnamefont{Newman}},
  \bibinfo{journal}{Proc. Natl. Acad. Sci. USA} \textbf{\bibinfo{volume}{98}},
  \bibinfo{pages}{404} (\bibinfo{year}{2001}).

\bibitem[{\citenamefont{Palla et~al.}(2008)\citenamefont{Palla, Farkas,
  Pollner, Der\'enyi, and Vicsek}}]{Pal_01}
\bibinfo{author}{\bibfnamefont{G.}~\bibnamefont{Palla}},
  \bibinfo{author}{\bibfnamefont{I.~J.} \bibnamefont{Farkas}},
  \bibinfo{author}{\bibfnamefont{P.}~\bibnamefont{Pollner}},
  \bibinfo{author}{\bibfnamefont{I.}~\bibnamefont{Der\'enyi}},
  \bibnamefont{and} \bibinfo{author}{\bibfnamefont{T.}~\bibnamefont{Vicsek}},
  \bibinfo{journal}{New J. Phys.} \textbf{\bibinfo{volume}{10}},
  \bibinfo{pages}{123026} (\bibinfo{year}{2008}).

\bibitem[{\citenamefont{Kitsak et~al.}(2010)\citenamefont{Kitsak, Gallos,
  Havlin, Liljeros, Muchnik, Stanley, and Makse}}]{Kit_01}
\bibinfo{author}{\bibfnamefont{M.}~\bibnamefont{Kitsak}},
  \bibinfo{author}{\bibfnamefont{L.~K.} \bibnamefont{Gallos}},
  \bibinfo{author}{\bibfnamefont{S.}~\bibnamefont{Havlin}},
  \bibinfo{author}{\bibfnamefont{F.}~\bibnamefont{Liljeros}},
  \bibinfo{author}{\bibfnamefont{L.}~\bibnamefont{Muchnik}},
  \bibinfo{author}{\bibfnamefont{H.~E.} \bibnamefont{Stanley}},
  \bibnamefont{and} \bibinfo{author}{\bibfnamefont{H.~A.} \bibnamefont{Makse}},
  \bibinfo{journal}{Nature Physics} \textbf{\bibinfo{volume}{6}},
  \bibinfo{pages}{888} (\bibinfo{year}{2010}).

\bibitem[{Exx({\natexlab{a}})}]{Exxon_05}
\bibinfo{note}{In Euclidean networks, the void or hole structure has been
  studied in many disciplines for example to characterize the morphology of the
  bones and to understand how habitat fragmentation affects animal movement
  processes.}

\bibitem[{\citenamefont{Leath}(1976)}]{Lea_01}
\bibinfo{author}{\bibfnamefont{P.}~\bibnamefont{Leath}},
  \bibinfo{journal}{Phys. Rev B} \textbf{\bibinfo{volume}{14}},
  \bibinfo{pages}{5046} (\bibinfo{year}{1976}).

\bibitem[{\citenamefont{Braunstein et~al.}(2007)\citenamefont{Braunstein, Wu,
  Chen, Buldyrev, Kalisky, Sreenivasan, Cohen, L{\'o}pez, Havlin, and
  Stanley}}]{Bra_01}
\bibinfo{author}{\bibfnamefont{L.~A.} \bibnamefont{Braunstein}},
  \bibinfo{author}{\bibfnamefont{Z.}~\bibnamefont{Wu}},
  \bibinfo{author}{\bibfnamefont{Y.}~\bibnamefont{Chen}},
  \bibinfo{author}{\bibfnamefont{S.~V.} \bibnamefont{Buldyrev}},
  \bibinfo{author}{\bibfnamefont{T.}~\bibnamefont{Kalisky}},
  \bibinfo{author}{\bibfnamefont{S.}~\bibnamefont{Sreenivasan}},
  \bibinfo{author}{\bibfnamefont{R.}~\bibnamefont{Cohen}},
  \bibinfo{author}{\bibfnamefont{E.}~\bibnamefont{L{\'o}pez}},
  \bibinfo{author}{\bibfnamefont{S.}~\bibnamefont{Havlin}}, \bibnamefont{and}
  \bibinfo{author}{\bibfnamefont{H.~E.} \bibnamefont{Stanley}},
  \bibinfo{journal}{I. J. Bifurcation and Chaos} \textbf{\bibinfo{volume}{17}},
  \bibinfo{pages}{2215} (\bibinfo{year}{2007}).

\bibitem[{Exx({\natexlab{b}})}]{Exxon_04}
\bibinfo{note}{Notice that $p$ is the fraction of links in a link percolation
  process, while $p^{v}$ is the fraction of void nodes for a void node
  percolation process.}

\bibitem[{\citenamefont{Stauffer and Aharony}(1985)}]{Sta_01}
\bibinfo{author}{\bibfnamefont{D.}~\bibnamefont{Stauffer}} \bibnamefont{and}
  \bibinfo{author}{\bibfnamefont{A.}~\bibnamefont{Aharony}},
  \emph{\bibinfo{title}{Introduction to percolation theory}}
  (\bibinfo{publisher}{Taylor \& Francis}, \bibinfo{year}{1985}).

\bibitem[{\citenamefont{Newman}(2005)}]{New_06}
\bibinfo{author}{\bibfnamefont{M.~E.~J.} \bibnamefont{Newman}},
  \bibinfo{journal}{Phys. Rev. Lett.} \textbf{\bibinfo{volume}{95}},
  \bibinfo{pages}{108701} (\bibinfo{year}{2005}).

\bibitem[{\citenamefont{Kenah and Robins}(2007)}]{Ken_01}
\bibinfo{author}{\bibfnamefont{E.}~\bibnamefont{Kenah}} \bibnamefont{and}
  \bibinfo{author}{\bibfnamefont{J.~M.} \bibnamefont{Robins}},
  \bibinfo{journal}{Phys. Rev. E} \textbf{\bibinfo{volume}{76}},
  \bibinfo{pages}{036113} (\bibinfo{year}{2007}).

\end{thebibliography}
\end{document}